# Modified Newtonian Dynamics, an Introductory Review

Riccardo Scarpa

*European Southern Observatory, Chile*
*E-mail rscarpa@eso.org*

**Abstract.** By the time, in 1937, the Swiss astronomer Zwicky measured the velocity dispersion of the Coma cluster of galaxies, astronomers somehow got acquainted with the idea that the universe is filled by some kind of dark matter. After almost a century of investigations, we have learned two things about dark matter, (i) it has to be non-baryonic -- that is, made of something new that interact with normal matter only by gravitation-- and, (ii) that its effects are observed in stellar systems when and only when their internal acceleration of gravity falls below a fix value $a_0=1.2\times10^{-8}$ cm s$^{-2}$. Being completely decoupled dark and normal matter can mix in any ratio to form the objects we see in the universe, and indeed observations show the relative content of dark matter to vary dramatically from object to object. This is in open contrast with point (ii). In fact, there is no reason why normal and dark matter should conspire to mix in just the right way for the mass discrepancy to appear always below a fixed acceleration. This systematic, more than anything else, tells us we might be facing a failure of the law of gravity in the weak field limit rather then the effects of dark matter. Thus, in an attempt to avoid the need for dark matter many modifications of the law of gravity have been proposed in the past decades. The most successful – and the only one that survived observational tests -- is the Modified Newtonian Dynamics. MOND posits a breakdown of Newton's law of gravity (or inertia) below $a_0$, after which the dependence with distance became linear. Despite many attempts, MOND resisted stubbornly to be falsified as an alternative to dark matter and succeeds in explaining the properties of an impressively large number of objects without invoking the presence of non-baryonic dark matter. This suggests MOND is telling us something important about gravity in the weak field limit. In this paper, I will review the basics of MOND and its ability to explain observations without the need of dark matter.

## INTRODUCTION

As far back as 1983, after realizing that mass discrepancies are observed in stellar systems *when and only when* the acceleration of gravity falls below a fix value[1], M. Milgrom proposed a modification of Newtonian dynamics (MOND) as an alternative to non-baryonic dark matter[2,3,4]. Milgrom's proposal states that gravity (or inertia) does not follow the prediction of Newtonian dynamics for acceleration smaller than $a_0=1.2\times10^{-8}$ cm s$^{-2}$, as derived from galaxy rotation curves fit[5]. Below this acceleration the behavior of gravity changes becoming asymptotically $a = \sqrt{a_N a_0}$, where $a_N$ is the usual Newtonian acceleration. The transition from Newtonian to MOND regime occurs for accelerations that remain undetected within the solar system where the strong field of the sun is dominating all dynamical processes. To be more precise, we see that even the acceleration by Mercury on Pluto is above $a_0$, so that the there is no way in the solar system to test the validity of Newtonian dynamics down to the acceleration regimes typical of galaxies. As a consequence, we have no direct proof of its validity below $a_0$, and its applicability in the realm of galaxies is not guaranteed.

Since the seminal papers by Milgrom[2,3,4] a few authors have worked on this subject showing how this simple -- though revolutionary -- idea explains many properties of galaxies without the need of non-baryonic dark matter ( e.g., [5,6,7,8]). MOND is so successful that, as a minimum, it is telling us the exact functional form of the force in galaxies. Any theory of galaxy and structure formation must therefore be able to reproduce the MOND phenomenology. MOND is also effective in describing the dynamics of galaxy groups and clusters[9,10] and, with some approximations, gravitational lensing[11,12]. For a recent review see [13].

As I shall show, MOND has been successfully tested in a very large number of cases and, were it not for dark matter being thoroughly ingrained in the mind of astronomers, by now there should be little doubt about its relevance in describing the universe. Nowadays, the problem is no longer to show whether MOND works -- because this was already done -- rather that MOND and dark matter give conceptually different but *operationally equivalent* description of cosmic phenomena. Because of this, astronomers have decided to stick to dark matter and continue to add more and more arbitrary entities, once called epicycles, to save the appearances.

To make progress in this field one has to find a way to show MOND - or equivalently dark matter -- effects in places where no one is expecting them. In the first place MOND should be tested in the laboratory (but see below Milgrom's opinion about this). Also, one can study the dynamics of very wide double stars, globular clusters or other small structure where the effects of dark matter are believed to be negligible. If this will be done --successfully of course -- in a sufficiently large number of cases, it will be no longer possible to claim that dark matter is the explanation and new physics will be required. However, before moving too far ahead, lets start comparing MOND predictions to real data.

## MOND BASICS

The MOND acceleration of gravity $a$ is related to the Newtonian acceleration $a_N$ by

$$a_N = a\mu(a/a_0).$$

The constant $a_0 = 1.2 \times 10^{-8}$ cm s$^{-2}$ is meant to be a new constant of physics.
The interpolation function $\mu(a/a_0)$ admits the asymptotic behavior $\mu=1$ for $a >> a_0$, so to retrieve the Newtonian expression in the strong field regime, and $\mu = a/a_0$ for $a << a_0$.

With these assumptions, the weak acceleration limit of gravity is:

$$a = \sqrt{a_N a_0} = \frac{\sqrt{GMa_0}}{r}, \qquad (1)$$

with dependence $1/r$ on distance $r$ from the body of mass M generating the field.

Different though functionally equivalent expression for $\mu$ have been used in the literature (e.g., [3,7,14]). Currently the most used function is:

$$\mu(a/a_0) = \frac{a/a_0}{\sqrt{1+(a/a_0)^2}}$$

which works well in describing galactic rotation curves and has the advantage of being simple. Solving for $a$ the MOND acceleration of gravity reads

$$a = a_N \left( \frac{1}{2} + \frac{1}{2}\sqrt{1 + \left(\frac{2a_0}{a_N}\right)^2} \right)^{1/2}$$

This expression allows the calculation of the MOND gravitational acceleration provided the Newtonian one is known.

## MOND AND GALAXY ROTATION CURVES

The functional form for MOND was chosen so to have a $1/r$ dependence of gravity at large radii. In this way flat rotation curves are automatically explained (Figure 1). However, galaxies have rotation curve of all forms, not necessarily flat at large radii. This is important to keep in mind. Every galaxy represents a truly independent test for MOND, in particular when the rotation curve is not flat (Fig. 2). When fitting rotation curves– note that $a_0$ is fixed and cannot change -- the only free parameter is the mass-to-light ratio, which is to a large extent fixed by stellar evolution theory. Thus even though sometime one has to allow the distance of the object to vary within reasonable values in order to achieve a decent fit, MOND gives very little flexibility. This is to be compared with the almost unlimited adjustability of the fit offered by the dark-matter, the perfect free parameter that can appear in any quantity and with any distribution. This fact is well exemplified in Fig 3 where an attempt is made to fit the rotation curve of a fake galaxy. This was obtained using the photometry of one object, which defines the mass distribution, and the rotation curve of another. Having no free parameters MOND *cannot* fit this curve, which is good, while dark matter model can easily fit this non-existing object, questioning the meaning of all the fits obtained with this method.

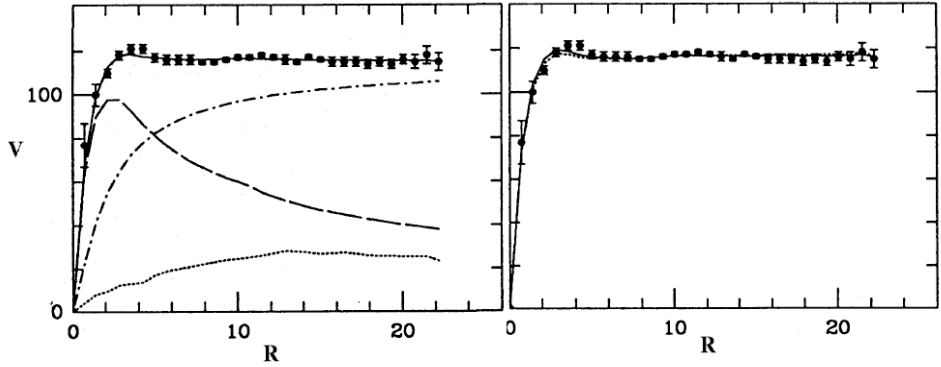

**FIGURE 1.** Example rotation curve of one normal high-surface-brightness galaxy (NGC 6503, from data published in [5]). **(Left)**: Three-parameter dark-halo fit (solid curve). The rotation curve of the stellar (dashed line), gas (dotted line), and dark-halo (dash-dotted line) components are shown. The fitting parameters are the M/L ratio of the disk, the halo core radius, and the halo asymptotic velocity. **(Right)**: The MOND fit showing the one-parameter fit (M/L). Velocities in km/s and radius in kpc.

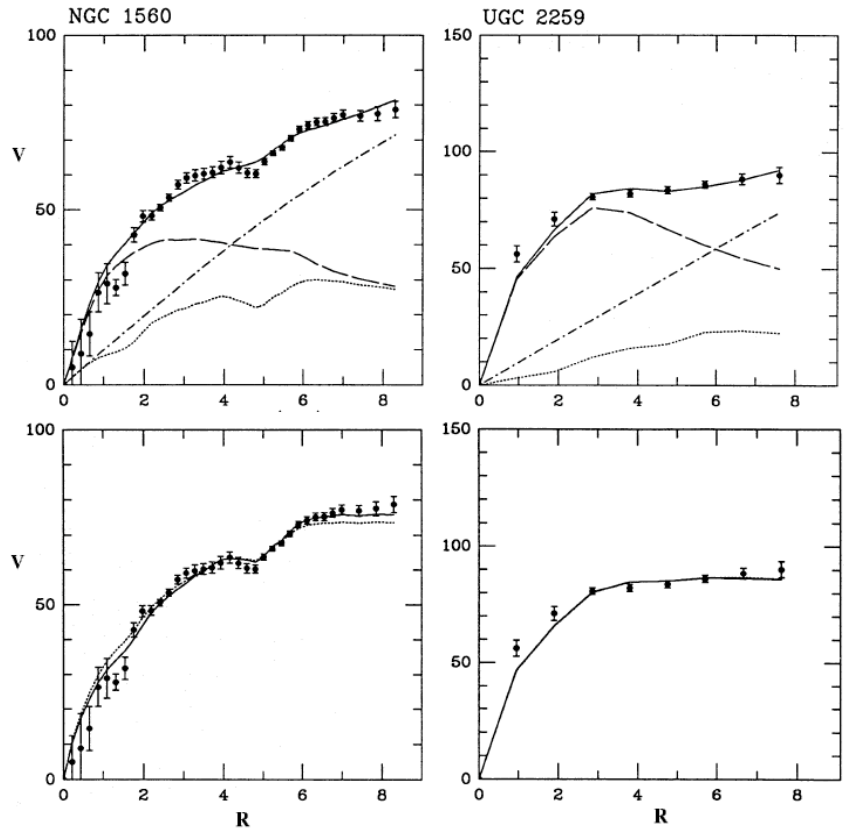

**FIGURE 2.** Rotation curves for two low surface brightness, gas-dominated galaxies (data from [5]). **Top**: Three-parameter dark-matter halo fits (solid curve). The rotation curve of the stellar (dashed line), gas (dotted line), and dark-halo (dash-dotted line) components are also shown. The fitting parameters are the M/L ratio of the disk, the halo core radius, and the halo asymptotic velocity. **Bottom**: the same rotaion curves as before, fitted with the MOND prescription. The dotted line shows the one-parameter (M/L) fit, while the solid line show the two-parameter (M/L and distance) fit. For UGC2259 the two lines coincide. The important thing here is that these galaxies are gas dominated so that M/L becomes irrelevant and the fit has no free parameters at all. Velocities in km/s and distances in kpc.

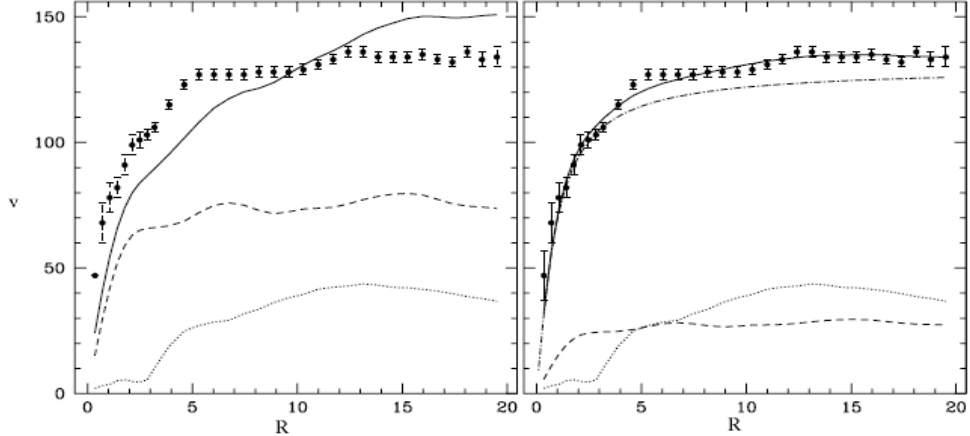

**FIGURE 3.** Fit of the rotation curves of an hybrid galaxy, obtained using the velocity information for one galaxy (NGC2403) and the photometry – used to compute masses – from another (UGC 128) (data from [15]). MOND (left panel) fails to fit this fake galaxy, which is good. On the other hand, a fit of an isothermal dark-matter halo (right panel) allows so much freedom that an acceptable fit can be found, which is bad. The curves relative to the gas component (dotted line) and stellar disk (dashed line) are also shown. Velocities in km/s, and radius in kpc.

## MOND AND THE T-F AND F-J RELATIONS

The Tully-Fisher relation[16] links the luminosity L of spiral galaxies to their asymptotic rotational velocity V in the form $L \propto V^4$. Similarly, the Faber-Jackson[17] relation, $L \propto \sigma^4$, links the luminosity of elliptical galaxies the their central velocity dispersion. Both relations find an immediate explanation within the contest of MOND because the mass is linked to the 4$^{th}$ power of velocity, not the square, and in absence of dark matter mass and light are proportional.

Indeed, equating the centripetal acceleration to the gravitational acceleration from eq. 1, and introducing the mass to light ratio $\tau = M/L$, we get:

$$v^4 = GMa_0 \propto \frac{M}{L}L = \tau L$$

From this relation we see that MOND explains the T-F relation provided $\tau$ is roughly constant in galaxies. This is certainly the case because all spiral galaxies share very similar stellar populations, as demonstrated by the similar colors and spectral properties. Moreover, the fact that $\tau$ is basically constant allows us to introduce one important *prediction* of MOND: the T-F relation is universal and all galaxies should follow the same relation.

To compare the MOND description of the T-F relation to the Newtonian one, we have to introduce the average surface brightness $\Sigma$ of galaxies, define as the luminosity divided by the surface of the galaxy. Assuming for the sake of simplicity that galaxies are round, we have $\Sigma = L/\pi r^2$. Then equating the centripetal acceleration to the acceleration of gravity we have:

$$v^4 = \frac{(GM)^2}{r^2} \propto \frac{M^2 \Sigma}{L^2} L = \tau^2 \Sigma L$$

Here we can see the T-F relation provided the product $\tau^2 \Sigma$ is constant. As we just pointed out $\tau$ is basically constant, thus we have to conclude that also $\Sigma$ is constant. This is certainly *not* the case because galaxies divide in two big families, the high surface brightness (HSB) and low surface brightness (LSB) galaxies. The latter having up to 100 times fainter surface brightness than normal HSB galaxies. Newtonian dynamics therefore predicts HSB and LSB galaxies should follow two *parallel but distinct* T-F relations. This is simply not the case (Figure 4). Both HSB and LSB galaxies follow the same T-F relation, as predicted by MOND. When Milgrom first made these considerations, no dynamical information were available for LSB galaxies, so this was a true prediction.

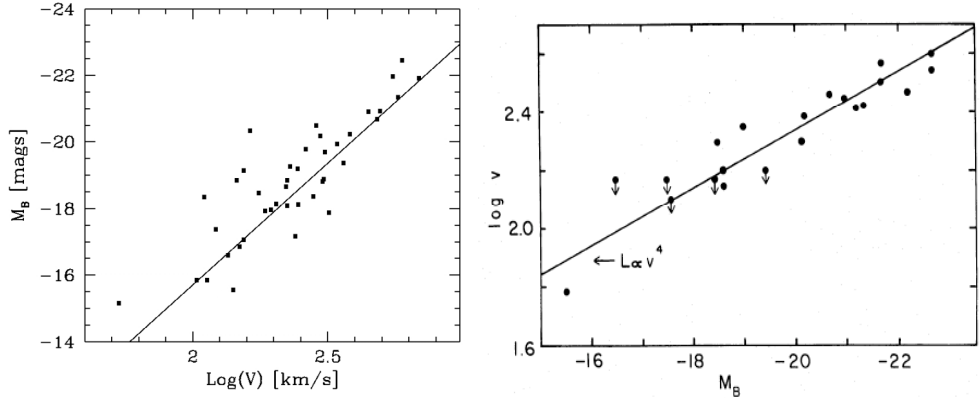

**FIGURE 4. (Left)** The Tully-Fisher relation for a sample of low surface brightness galaxies (points) is compared to the relation defined by HSB galaxies (solid line). Data refers to the two homogeneous datasets presented in [18]. As predicted by MOND and contrary to the Newtonian prediction, HSB and LSB galaxies follow the same relation. **(Right)** The Faber Jackson relation for elliptical galaxies links the central velocity dispersion to the total luminosity. The line correspond to $L/L_\Theta = 10\sigma^4$ (with $\sigma$ in km/s). Reproduced from [17].

## MOND AND ELLIPTICAL GALAXIES

Due to the high central density a considerable fraction of elliptical galaxies is in Newtonian regime of accelerations. This is why the properties of elliptical galaxies can be explained to a large extent without the needs for dark matter. This also means that in first approximation the virial theorem applies. Velocity dispersion $\sigma$, size, and mass of elliptical galaxies are therefore linked and when represented in the space defined by these three variables, they should define a plane. Mass, however, is never directly observed so that to verify whether this is true, one has to find a way to estimate it. Within the contest of MOND this is trivial, all we need is to convert the observed light into a mass via $\tau$, which is sensible constant among ellipticals. Alas, MOND predicts that when represented in the space defined by luminosity, size, and velocity dispersion all elliptical galaxies should follow a plane (a surface to be precise, as we will se in a moment).

By contrast, if we believe galaxies are filled by huge amounts of dark matter, distributed in arbitrary ways, there is no relation between luminosity and mass. The definition of *size* also become fuzzy, because dark matter is much more extended than the visible galaxy – indeed the "end" of dark haloes was never observed so far -- that one can not use the observed radius in the virial formula. As a consequence, elliptical galaxies *should not* follow a plane in this particular space, something in open contrast with the fact that elliptical galaxies are known to define a plane -- the "fundamental plane"-- in full agreement with MOND prediction.

To explain the existence of the fundamental plane (FP) one has to assume that dark and luminous matter are mixed in a constant ratio in elliptical galaxies. Even if this were true, however, things are not fine because the fundamental plane is tilted with respect to the plane defined by the virial relation (Fig. 5).

Finding an explanation for the tilt has been and still is one of the major puzzles in modern astrophysics. In a nutshell, both possibilities of a systematic change of the mass-to-light ratio $\tau$ with luminosity and departure from homology were fully investigated. To preserve the narrowness of the FP, the tilt requires dark matter to be finely tuned to the galaxy luminosity, or the departures from homology to be finely tuned to the galaxy size[19]. Of the two options the former is the most widely accepted, and the tilt ascribed to $\tau \propto L^{0.25}$ (e.g., [20]). Whatever the reason for the tilt is, both these possibilities face paramount difficulties as soon as LSB elliptical galaxies are considered, because these galaxies define a completely different plane[21,22] implying $\tau \propto L^{-0.4}$.

As noted before, in absence of dark matter Newtonian dynamics predicts galaxies to follow a perfect plane. This is only approximately true for MOND. The central region of elliptical galaxies is in Newtonian regime while the outskirt is not. Thus, according to the fraction of the galaxy that is in MOND regime – remember the effective radius used to place galaxies in the plane contains always half the total light -- galaxies deviate more or less significantly from the virial relation. How important MOND effects are on galaxies of different size/type is determined by their average surface brightness (Fig. 6). This is seen to vary in a systematic way. HSB galaxies became progressively *less* dense as they increase in size or luminosity. For them the MOND effects are therefore increasingly relevant and deviation from the virial relation becomes progressively more evident as we move from small to large galaxies. On the other hand, it is seen that LSB galaxies become progressively *more* dense as their size increase, and MOND effect get smaller. This is exactly what is observed, and the trend defined by MOND fully agrees with the measure slopes (Figure 7).

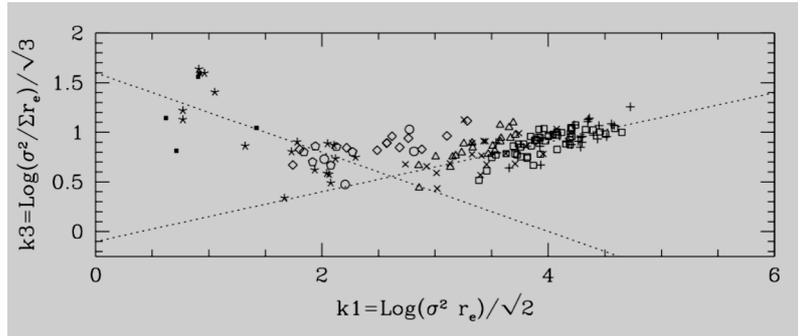

**FIGURE 5.** Edge on view of the fundamental plane of elliptical galaxies. The quantities k1 and k3 as defined in [19] are used. In Newtonian dynamics, $k_1$ is proportional to the total galaxy mass, while $k_3$ is proportional to $\tau$. The two dotted lines with slope -0.4 and +0.25, show the dependence of $\tau$ on M for LSB and HSB, respectively. Data were taken from [19]: Squares= giant ellipticals; Triangles= intermediate ellipticals; Diamonds= dwarf ellipticals; circle= compact ellipticals; Crosses=bulges; Dots= LSB ellipticals; Stars= LSB elliptical from [22]; Pentagons= dwarf elliptical in Virgo from [23]; Pluses= radio galaxies from [24].

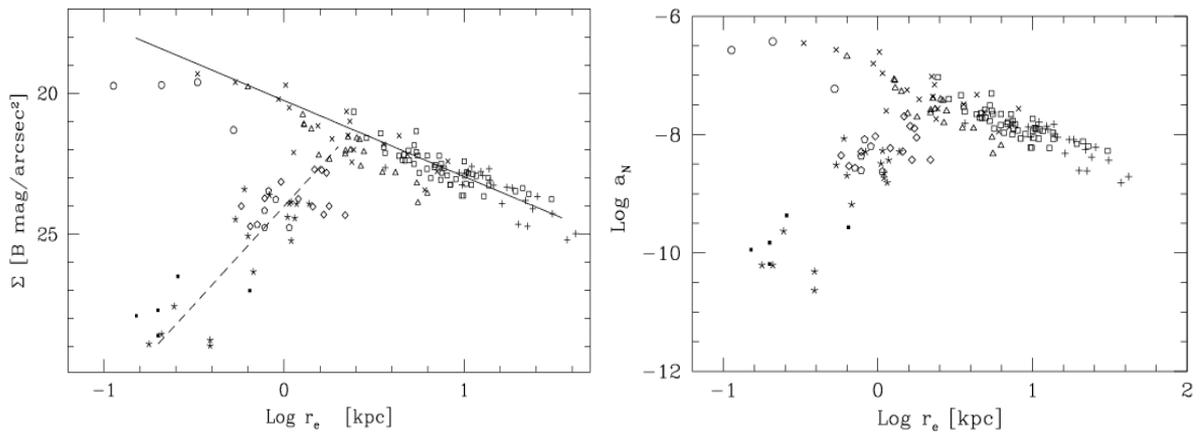

**FIGURE 6. (Left)**: The Kormendy relation showing the trends of the mean surface brightness $\Sigma$ as a function of the effective radius for HSB and LSB elliptical galaxies. HSB galaxies becomes progressively more diffuse with size, while LSB galaxies do exactly the opposite, becoming more dense as their size increase. **(Right)**: The trends in surface brightness are just trends in internal acceleration of gravity $a_N$. The different trends of the mass-to-light ratio $\tau$ observed in HSB and LSB galaxies can be understood in terms of the different strength of the MOND effects. Symbols as in Figure 5.

## MOND AND ULTRA COMPACT DWARF GALAXIES

Dwarf galaxies have attracted the attention of astrophysicists in the last years for many reasons. Among others, these galaxies are found to contain an impressive amount of non-baryonic dark matter[25,26] having mass-to-light ratios up to 100. The discovery[27,28,29] in an all-object survey of the Fornax cluster of a new type of galaxies, the ultra-compact dwarfs (UCD), has added a new member to the dwarf family. Unlike other dwarfs, however, UCD galaxies have very high central concentration and thus star-like morphology in typical one-arcsecond-resolution ground-based imaging. At the distance of the Fornax cluster (20 Mpc; Hubble constant $H_0 = 75$ km s$^{-1}$ Mpc$^{-1}$) this implies sizes of 100 pc at most. The B band luminosity ranges from to $3\times10^7$ $L_\odot$, significantly brighter than the brightest globular cluster known. Thus, UCDs are squarely placed midway between regular dwarf galaxies and globular clusters.

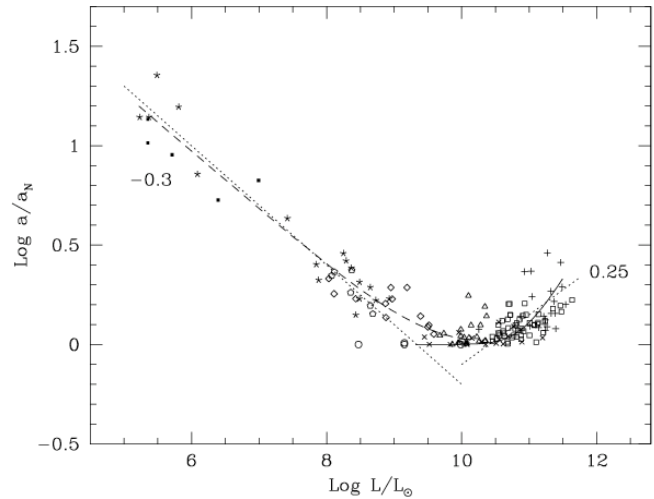

**Figure 7.** Plot of $a/a_N = \tau/\tau_0$ versus luminosity (symbols as in Figure 5). The meaning of this plot can be understood noticing that the discrepancy in acceleration is interpreted as due to extra mass or, in other words, as a variation of the mass-to-light ratio $\tau$ with respect to the true value $\tau_0$. The mass discrepancy decrease (increase) with luminosity for LSB (HSB) galaxies. The dependence is indicated by two dotted lines, with slope $\tau \propto L^{-0.3}$ for LSB and $\tau \propto L^{0.25}$ for HSB galaxies, respectively. In absence of dark matter $k1 \propto L$, thus this plot is equivalent to Fig 5. The observed slopes for $a/a_N = \tau/\tau_0$ are teh explanation of the tilt in the framework of MOND. The dashed and solid lines are the same drawn in Fig. 6 and are meant to visualize the loci followed by LSB and HSB galaxies.

Recent determination[30] of their effective radius (ranging from 10 to 22 pc) and velocity dispersion (ranging from 24 to 37 km s$^{-1}$) indicates that, if any, there is very little dark matter in UCD galaxies. Indeed assuming UCDs are virialized structure, masses ranging from $10^7$ to $10^8$ $M_\odot$ are found[30]. The corresponding $\tau$ varies from 2 to 4 in solar units, fully consistent with the expectation for an old stellar population without significant amount of dark matter. In the literature two different scenarios have been proposed to explain why UCDs show no mass discrepancy. One possibility is that UCDs are giant globular clusters[31] possibly the result of the merging of the giant stellar clusters created during periods of strong galaxy interaction[32], like the one observed in the ``Antennae'' system (e.g., [33]). The other possibility is suggested by the fact that UCDs luminosity and size match well the one of the *nucleus* of dwarf galaxies. It is therefore possible that UCD galaxies are the remnant of normal nucleated dwarf galaxies that have lost their external halo and dark matter, because of strong and repeated interaction with other galaxies. This latter scenario, referred to as galaxy threshing[34], is the most widely accepted.

Thinking in terms of MOND, the word *compact* is sufficient to know that no mass discrepancy should be found in these objects. To show that is the case, lets put a lower limit to the internal acceleration of gravity taking the less luminous UCD having luminosity $4\times10^6$ $L_\odot$, assigning to it the maximum possible diameter 100 pc, and the minimum mass-to-light ration ($\tau=2$). At the outer edge of this galaxy the acceleration is $2\times10^{-8}$ cm s$^{-2} > a_0$. According to MOND, then, UCDs are everywhere in Newtonian regime and *no non-baryonic dark matter is to be found*. Up to now, MOND is in full agreement with observations.

It is also clear that the reason why we do not observe a mass discrepancy is neither due to the particular evolutionary history of this galaxies, nor to the fact that UCDs might be giant globular clusters. It is solely due to the fact that they are in Newtonian regime and therefore should strictly obey Newtonian dynamics.

## MOND AND GALAXY CLUSTERS

Moving to cluster of galaxies we see that also for these structures MOND gives reasonable result[10]. However, rich galaxy clusters do represent the only objects where the agreement is not as good as we have seen so far. It is evident from Figure 8 that the mass discrepancy does not completely disappear, there is a remaining systematic discrepancy between the MOND dynamical mass and the observed mass. How severe this is for MOND is not clear. Certainly it is at least a good sign that MOND masses are *larger* that the total mass observed, indicating some *baryonic* component of the clusters still remains undetected. If the discrepancy were the other way around, it would be devastating for MOND.

In many astrophysical contexts, a factor of 2 discrepancy can often be accommodated by reconsidering the effects of the several idealized assumptions that are not realized in every case. A fair number of simplifying assumptions is certainly necessary in order to be able to derive cluster masses, so it would not be surprising if this discrepancy will disappear in the future.

However it is quite possible, and perhaps even good for MOND, that the residual mass discrepancy is real and that the accounting of baryons in clusters is not yet complete. Indeed, this seems to be required, even in the context of MOND, by strong gravitational lensing.

The critical surface density required for strong lensing is

$$\Sigma_c = \frac{1}{4\pi} \frac{cH_0}{G} F$$

where $F \sim 10$ is a dimensionless function of the lens and source redshifts[35], c is the speed of light, and $H_0$ the Hubble constant. MOND applies at surface densities[2] below $\Sigma \sim a_0/G \sim \Sigma_c/5$, that is to say, the critical surface density for strong lensing is always greater than the upper limit for MOND phenomenology. *Strong lensing never occurs in the MOND regime*. Strong lensing observed in clusters typically requires a total projected mass in the inner 100–200 kpc between $10^{13}$ and $10^{14}$ $M_\Theta$, which is not present in the form of hot gas or stars. Whether or not MOND is correct, undetected matter does seem to be required in the cores of rich clusters. Interestingly, masses of this order of magnitude would make the residual discrepancy disappear. Thus one can imagine the MOND value of the dynamical mass is correct, and a prediction is made that the remaining *baryonic* mass to be found in cluster is within a factor 2 of the currently observed mass.

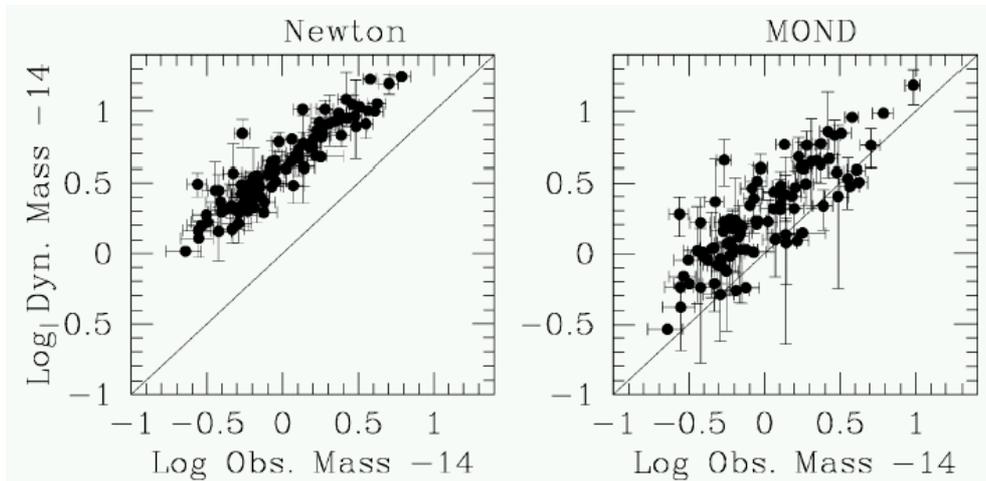

**FIGURE 8. (Left)** The Newtonian dynamical mass for galaxy clusters plotted against the total observable mass (gas+stars). The solid line represents the locus of no mass discrepancy. **(Right)** The factor ten discrepancy is reduced by MOND down to a factor two. Reproduced from [10].

# MOND AND GRAVITATIONAL LENSING

As seen in the previous paragraph a strong gravitational field is needed to significantly deflect light and produce multi-images of a background source. Thus strong gravitational lensing always occurs in Newtonian regime and do not constitute a test for MOND.

Weak lensing, however, do occur in the weak field regime and we can use the weak tangential alignment, the cosmic shear, of distant galaxies caused by nearby ones to test MOND.

To do the test we need to compute the effects of MOND on light. In general relativity a photon experiences twice the deflection of a massive particle moving at the speed of light. Thus, as normally done in Newtonian dynamics, in first approximation in the contest of MOND one has to compute the deviation for a massive particle and than double the effect. This is what was done so far to test MOND.

Considering the simplest case of a test particle that does not disturb the external mass distribution, and with speed such that its path is nearly linear, the deflection can be approximated by integrating the acceleration along the unperturbed trajectory (i.e. a straight line). Further restricting to the case of a point mass M, the asymptotic deviation angle (in radians) is

$$A_M = \frac{2\pi}{c^2}\sqrt{GMa_0}.$$

The deflection of light at large impact parameters from a point mass in the MOND framework is independent from the impact parameter, as much as the rotational velocity is independent from distance. This can be easily understood because the MOND acceleration at large impact parameters is proportional to 1/r, while the period of time over which this acceleration applies is proportional to r. In this way the deflection of light at increasingly large impact parameters is constant (Figure 9).
The deflection angle is not directly measurable, but the distortion of images – the shear -- and the (total) magnification of sources can be calculated directly from $A_M$, and these are observable.

So far the cosmic shear has been used to derive the dark matter content of the deflectors (e.g. [36]). Not surprisingly, results are all consistent with galaxies being surrounded by very extended and massive isothermal haloes. Interestingly, no upper limit has been placed on the halo size, despite the fact that a systematic distortion has been measured out to hundreds of times the visible extent of the foreground galaxies. This is by far the most stringent lower limit on the size of dark matter haloes that, to all practical effects, seems to be infinite.
MOND results are once again consistent with observations (Figure 10).

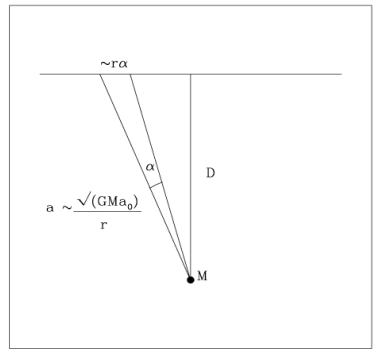

**FIGURE 9.** The deflection of light at large impact parameters D is constant because $a \propto 1/r$ while the time over which this acceleration applies is dt=rdα/c∝r. The net effect is that the deviation of light depends only on the mass M of the object.

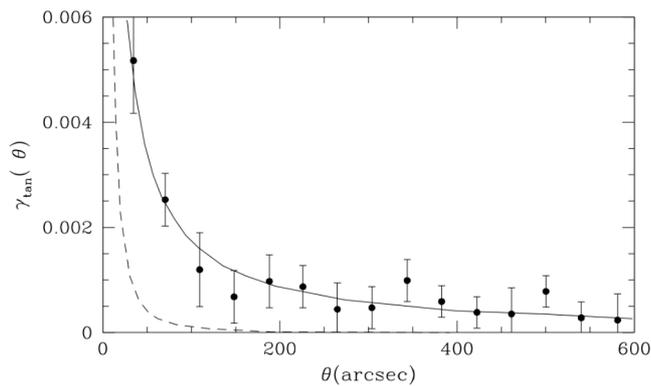

**FIGURE 10.** The mean cosmic shear γ around foreground galaxies compared to Newtonian and MOND prediction. The original data [37] in the g', r', and i' photometric bands have been averaged, excluding the first and third g' points that clearly deviated from the values found in the other two bands The models shown are the one presented in [38] and refers to the MOND prediction (solid line), and the Newtonian result if there is no dark matter (dashed line).

# MOND AND THE BACK HOLE – BULGE MASSES RELATION

As soon as the mass ($M_{BH}$) of a number of black hole located at the center of galaxies were measured it was recognized that $M_{BH}$ is related to the global properties of the galaxy. First a correlation was noted[39] between $M_{BH}$ and the luminosity of the spheroidal component -- the bulge – of the form $M_{BH}/M_\Theta=0.01L_B/L_\Theta$ (Figure 11). Later a possibly tighter relation $M_{BH}/M_\Theta=0.07\sigma^4$ was found between $M_{BH}$ and velocity dispersion $\sigma$ of the bulge[40, 41]. The physical processes responsible for such correlations remain unclear, though they are important because they help understanding the BH-galaxy formation process, also offering a tool to estimate BH masses by the much easier direct observation of the global properties of the host galaxy. The problem here is that in the presence of dark matter it is difficult to reconcile the two relations. While $\sigma$ is somehow proportional to the bulge mass (including dark matter), the bulge luminosity $L_{Bulge}$ is certainly not. Nevertheless, the two relations are necessarily linked because both are pointing to the same thing, the BH mass. Because of this inconsistency, the two major groups involved in the debate soon started arguing whether only the $M_{BH}$-$\sigma$ relation should be considered and the $M_{BH}$-$L_{Bulge}$, due to the somewhat larger scatter, neglected. A "war" also ignited around the correct slopes for both relations and up to now no agreement was reached.

By contrast, according to MOND Nature has chosen the simplest solution (always preferable): every bulge puts a fix fraction of its total mass into its central BH.

Indeed, in *absence* of dark matter, we have $M_{BH}/M_\Theta=0.01L_B/L_\Theta =0.005M_{Bulge}/M_\Theta$ (for $\tau=2$). Similarly, keeping in mind the F-J relation (Figure 4) we have $M_{BH}/M_\Theta=0.07\sigma^4 = 0.007L_{Bulge}/L_\Theta= 0.0035M_{Bulge}/M_\Theta$ fully consistent with the previous one.

MOND here is relevant in telling us that the $M_{BH}$-$L_{Bulge}$ has slope exactly 1, and that the right exponent for the $M_{BH}$-$\sigma$ relation is exactly 4, value that just happens to be the average of the two values proposed by the two major groups involved in this field.

Thinking out of the box, one can point out that MOND implies the F-J relation to be universal for spheroids, meaning also LSB galaxies should follow the same relation for BH mass. At present it is not at all clear whether such diffuse objects can form a central massive object so it will be interesting to make experiments to see whether this is the case and, if yes, how BH masses compare with this prediction.

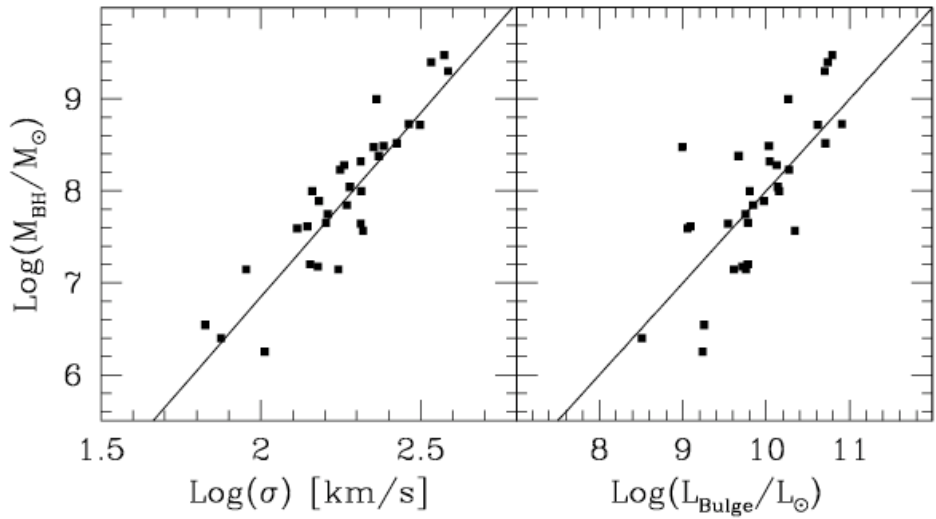

**FIGURE 11. (Left)** The black hole mass – bulge central velocity dispersion relation is well described by the relation $M_{BH}/M_\Theta=0.07\sigma^4$ (solid line, $\sigma$ in km/s). **(Right)** The bulge luminosity – black hole mass relation is consistent with the linear relation $M_{BH}/M_\Theta=0.01L_B/L_\Theta$. Both relations are well understood in absence of dark matter. Data are from the compilation of [42].

## MOND AND GLOBULAR CLUSTERS

Of the many possible tests for MOND the most effective are the one involving the study of objects believed to be free from the effects of dark matter. The only test of this type so far attempted involves globular clusters. It is well know that the properties of these objects can be explained *without* resorting to dark matter. Thus these objects should follow Newtonian dynamics down to whatever weak acceleration, unless Newton's law breaks down below some acceleration.

In the framework of MOND globular clusters agree to Newtonian dynamics because they are extremely compact and the internal gravitational field strong. Moving far enough from their center, however, MOND predicts the field will eventually deviate from the Newtonian one. If this is the case, then it will be difficult to explain the deviations specifically because dark matter cannot be invoked. A first pilot experiment for the globular cluster Omega Centauri was presented in [43] and a full account of the status of these experiments is given elsewhere in this book (Scarpa, Marconi & Gilmozzi). Here, it should suffice to say that globular clusters *do show* MOND effects as soon as their internal gravitational field goes below $a_0$, fully confirming the MOND phenomenology also in these class of objects and perhaps representing the ultimate test against the existence of non-baryonic dark matter.

## CONCLUSIONS

It should be clear by now that wherever the acceleration of gravity goes below $a_0$ we should detect MOND effects. This general statement, combined with a very simple prescription for computing gravitational accelerations in the weak filed limit, is extremely effective in describing the properties of many different objects seen in the universe. For instance, lets have a look at Figure 12 where dynamical and luminous masses are compared for about 1000 objects. In this figure we see how precisely MOND eliminates the mass discrepancy in objects ranging from globular clusters to clusters of galaxies that are at least $10^4$ times bigger and $10^7$ times more massive. Not only, it also explains why objects as diverse as LSB dwarf galaxies and group of galaxies do show the same mass discrepancy inspite of the vastly different physical properties. A particularly interesting case is the one of the globular cluster Palomar 13 that has been reported to have unusually large velocity dispersion. This is a very diffuse cluster with internal accelerations of the order of $10^{-9}$ cm s$^{-2}$ that looks like an open cluster but is as old as globular clusters. Well, Palomar 13 do follow the MOND law and show the same mass discrepancy of galaxies of comparable surface brightness, as is clearly seen in Fig. 12.

Looking at this figure, one has the feeling that MOND applies universally. Milgrom, however, has repeatedly stated the MOND formalism applies only to objects for which the *total* gravitational field in below $a_0$. That is to say, in determining whether MOND effects are relevant for a given object, any external contribution from nearby objects must be taken into account. The reasons for this claim rest on the fact that nearby open stellar clusters – which experience a field of the order of $a_0$ due to the Milky Way while having a very weak internal field -- do not show large mass discrepancies. To explain this apparent inconsistency, Milgrom was led to conclude that what is relevant is the total field. All tests *but one* done so far do satisfy this restriction (total field below $a_0$). We therefore are not in the position to tell whether MOND applies only when the total acceleration is below $a_0$ or if the breakdown of the Newtonian dynamics occurs irrespectively of the external field. According to Milgrom's point of view, MOND cannot be tested within the solar system and one as to move as far as 0.1 light years from the sun before MOND effects could be detected. Thus, depending on the proximity to massive objects, an object normally showing large mass discrepancies should follow the Newtonian dynamics below $a_0$. A good example of this situation is the case of globular clusters orbiting nearby the Milky Way. Up to now MOND was tested in three globular clusters one of which, NGC 6171, was studied precisely for being close enough to the Galaxy that the total field determining the dynamics of its external regions is above $a_0$. Nevertheless, this cluster seems to follow the MOND law. At the time of writing this result, being limited to one object, is not conclusive so we are not in the position of drawing any firm conclusion. More data are being collected at the European Southern Observatory VLT telescope in order to clarify the issue.

It is worth to point out that open clusters *are not* in dynamical equilibrium and are known to disrupt rather quickly. For instance, the vast majority of open clusters are not able to survive even for a single complete rotation of their own host galaxy. Their internal velocity dispersion, therefore, is not representative of their total mass and some other effect, yet to be understood, are responsible for it. In my opinion, open clusters put no constraints on MOND and the claim by Milgrom is unjustified.

Thus, it is my personal opinion – and I am the only one responsible for it if proved wrong-- that if Newtonian dynamics fails below $a_0$, this should be true irrespectively of the total field and one should be able to observe MOND effect also here on earth. For instance, I think a refined version of the Cavendish experiment studying gravitational forces in the horizontal plane should detect MOND effects.

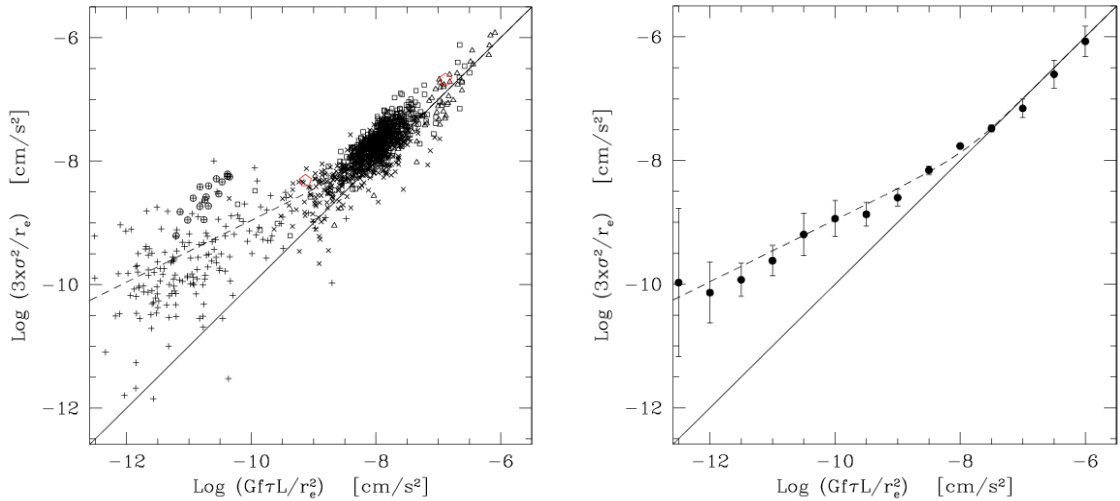

Figure 12. Comparison of the *Newtonian* acceleration of gravity computed using dynamical masses, versus luminous masses (i.e., converting light into mass via τ) for about 1000 objects. Symbols as in Figure 5 (**Left**) In the upper left corner are globular clusters (triangles) for which no mass discrepancy is observed. Moving to galaxies (squares and crosses) we start seeing deviation that become evident when considering LSB galaxies. Groups of galaxies (pluses) do show an enormous mass discrepancy. Rich clusters (circles with pluses) are the only class of objects that seem to deviate systematically from MOND prediction. Finally, two large pentagons represent the globular cluster Omega Centauri and Palomar 13. The solid line shows the locus of objects for which dynamical and luminous masses agree while the MOND value of the acceleration -- computed from luminous masses – is shown by the dashed line. (**Right**) Due to the inevitable dispersion of the data, the agreement between MOND and observations became more evident after rebinning. It is seen that MOND agrees with observations over at least 6 orders of magnitude in acceleration.

## ACKNOWLEDGMENTS

I am grateful to the dr. Eric Lerner and dr. Jose Almeida for making this conference possible and for giving me the possibility to freely express my idea, something that would have not be possible in other forums.